\documentstyle[preprint,epsf,prl,aps]{revtex}
\tightenlines
\begin{document}
\draft
%@twocolumnfalse\endcsname

\title{Large deformation of spherical vesicle studied by 
perturbation theory and {\bf \it Surface Evolver} }
\author{Jianjun Zhou$^{1}$, Yong Zhang$^{1}$, Xin Zhou$^{1}$ and
  Ou-Yang Zhong-can$^{1,2}$ }
\address{$^{1}$Institute of Theoretical Physics,
 The Chinese Academy of Science,
 P.O.Box 2735 Beijing 100080, China}
\address{$^{2}$Center for Advanced Study,
 Tsinghua University, Beijing 100084, China}
\maketitle         
\begin{abstract}
%%%%%%%%%%%%%%%%%%%%%%%%%%%%%%%%%%%%%%%%%%%%%%%%%%%%%%%%%
With tangent angle perturbation approach the axial symmetry deformation
of a spherical vesicle in large under the pressure changes is studied 
by the elasticity theory of Helfrich spontaneous curvature model.  
Three main results in axial symmetry shape:
biconcave shape, peanut shape, and one type of
  myelin are obtained. These axial
symmetry morphology deformations are in agreement with those
  observed in
 lipsome experiments by dark-field 
light microscopy [Hotani, J. Mol. Biol. {\bf 178}, 113 (1984)]
 and in the red blood cell with two thin filaments (myelin) 
observed in living state (see,   
Bessis,  {\it Living Blood Cells and Their Ultrastructure}, 
Springer-Verlag,  
1973). Furthermore, the biconcave shape and peanut
shape can be simulated with the help of a powerful software,
Surface Evolver [Brakke, Exp. Math. {\bf{1}}, 141 (1992)], 
in which the spontaneous curvature can be easy taken
into account.

%%%%%%%%%%%%%%%%%%%%%%%%%%%%%%%%%%%%%%%%%%%%%%%%%%%%%%%%%
%%%%%%%%%%%%%%%%%%%%%%%%%%%%%%%%%%%%%%%%%%%%%%%%%%%%%%%%%%%%
\end{abstract}
%\pacs{PACS numbers:87.16.Dg,47.10.+g,68.15.+e,02.40.Hw}
%\vskip2pc]
%\newpage
\section{introduction}

Since the discovery of optical microscope, people have recognized that the 
human red blood cells (RBCs) \cite{Bessis} and lipsomes \cite{Hotani} can be in many beautiful and strange shapes
depending on the environment in which they exist, such as  pH value, osmotic pressure difference.
 The simplest model to explain these shapes is viewing 
RBC and lipsome as a vesicle of amphiphile bilayers. 

For a long time, both the experimental and the theoretical aspects of 
amphiphile bilayers and monolayers have gathered
much attention from physicists and chemists \cite{review}. 
Helfrich has shown the importance of spontaneous
curvature and has developed a phenomenological theory for the elasticity 
of fluid membranes by an analogy with 
the curvature elasticity of liquid crystals \cite{$Helfrich_1$}. Many authors
 have studied the 
deformation of vesicles with different shapes
 among which the spherical vesicle plays a special role in these 
studies. 
 The general shape equation of vesicles has been derived with 
differential geometry method  \cite{$Ou-Yang_1$}, then 
some interesting numerical  and analytical results  
have
been reported \cite{Deuling,Naito}. In the equation, the
osmotic  pressure  difference 
between the external and internal sides of the vesicle membrane,
  especially the spontaneous curvature, and elastic modulus are important quantities. Among them, 
 the influence of pressure on morphologic appearance of vesicle is
the most important one to be 
considered. In shape study, RBC becomes an attractive model of 
vesicle in living cells. On the other hand, Hotani  reported the lipsome shape transformation pathway
when it was put in solution at certain concentration, and pointed out that
osmotic pressure was found to be the driving force for the sequential
transformations \cite{Hotani}. In the experiment the concentration 
changes slowly,
and so does pressure, therefore, the shape formation of the lipsome can be seen as an equilibrium
 problem.  In Hotani's observation, single sphere finally formed 
and seemed to be stable forms because they showed no further morphological
change. Jenkins and Peterson  studied in detail the
 problem 
of the stability of RBC shapes \cite{Jenkins}, in 
which spherical vesicle was limit
case. Ou-Yang and Helfrich  studied instability and deformation
of a spherical vesicle by pressure, and they found that any infinitesimal
deformation corresponding to spherical harmonics $Y_{lm}$
would require a pressure difference larger than some threshold values 
\cite{$Ou-Yang$}.
The nonaxisymmetric deformation of  RBC is investigated with a contour
perturbation approach \cite{$Ou-Yang_2$}. Obviously, the above 
approach is not able to deal with a large deformation of a vesicle shape
, such as the myelin form in RBCs \cite{Bessis}. It then 
becomes a challenge to search for a method by which the large 
deformation of a spherical vesicle can be treated in theory of
perturbation. In the present work we report an useful issue on
solving the problem. The axial symmetry vesicle contour is expressed as 
$Z=-\int \tan \Psi d \rho$ where $\Psi$ is the tangent angle of the 
contour (see Fig.~\ref{1}).
 One can find that a small change of $\Psi$ at $\Psi$ 
$\rightarrow$ $\pi/2$ is
able to cause a large deformation in contour. 
Just like the method of Deuling and Helfrich \cite{Deuling}, we
introduce an alternative variable of $\sin \Psi$, $u$, thus, the
contour Z will change with considerable amplitude as $\sin \Psi$
has small perturbation. with the issue we can calculate 
the large deformation of the contour via perturbation approach. 

 In this paper, we do some calculations
 about the large deformation of a 
spherical vesicle with the new approach via pressure perturbation, varied biconcave
shapes and one kind of peanut shape can be obtained. Especially we 
show one type of myelin form with two long filaments attaching on the spherical vesicle.
To confirm the analytic 
calculation we also perform some computer simulations with the help
of a powerful software, {\it Surface Evolver} \cite{Brakke}. 
%%%%%%%%%%%%%%%%%%%%%%%%%%%%
%%%%%%%%%%%%%%%%%%%%%%

\section{perturbation theory}
Equilibrium shapes of phospholipid vesicles are assumed to correspond to the 
local minima of the elastic free energy of
these systems. In the Helfrich spontaneous curvature model, the  
spontaneous curvature plays a fundamental
role in accounting for the different morphologic appearances of 
vesicles \cite{$Helfrich_1$}. According to Helfrich's theory,
the  free energy of a vesicle is written as 
\begin{eqnarray}
F=\frac{1}{2} k_c \oint (c_1+c_2-c_0)^2 dA +\Delta p\int dV + 
\lambda \oint dA,  \label{1}
\end{eqnarray}
%%%%%%%%%%%%%%%%%%%%%%%%%%%%%%%%%%%%%%%%%%%%%%%%%%%%%%%%
where $k_c$ is the bending elastic modulus;  $c_1$ and $c_2$ are
 the two principle curvatures of the surface of the vesicle,  $c_0$ is the spontaneous
curvature to describe the possible asymmetry of the outer and inner layers
of membrane, and  dA and dV are the surface area and the volume elements
for the vesicle, respectively. 
 The Lagrange Multipliers $\Delta p$ and 
$\lambda$ take account of the constraints of constant volume and area, 
respectively, and can be physically understood as the
pressure difference between the external and the internal environment and 
the surface tensile coefficient. 
     The general shape equation has been derived via variational 
     calculus \cite {$Ou-Yang_1$} to be
\begin{eqnarray}
&\Delta& p - 2 \lambda H + k_c (2 H+c_0) (2 H^2-2 K-c_0 H)+
2 k_c \bigtriangledown^2 H \nonumber\\
&&=0, \label{2}
\end{eqnarray}
where $H$ and $K$ are the mean and the Gaussian curvatures, respectively, 
and $\bigtriangledown^2$ is the Laplace-Beltrami 
operator \cite{$Ou-Yang$}.
  Assuming that the shape has axial symmetry, the general shape equation 
  becomes a third order nonlinear 
differential equation
%%%%%%%%%%%%%%%%%%%%%%%%%%%%%%%
\begin{eqnarray}
&\cos^3& \Psi \frac{d^3\Psi}{d\rho^3}= 4\sin \Psi \cos^2 \Psi  \frac{d^2\Psi}{d\rho^2}\frac{d\Psi}{d\rho} \nonumber \\ 
&&-\cos\Psi (\sin^2\Psi-\frac{1}{2} \cos^2 \Psi) (\frac{d\Psi}{d\rho})^3 
\nonumber\\
&&+7 \frac{\sin \Psi \cos^2 \Psi}{2 \rho} (\frac{d\Psi}{d\rho})^2-
\frac{2 \cos^3 \Psi}{\rho} (\frac{d^2\Psi}{d\rho^2}) \nonumber\\
&&+\left[\frac{c_0^2}{2}-\frac{2 c_0\sin\Psi}{\rho}+\frac{\lambda}{k_c} 
-\frac{\sin^2\Psi-2 \cos^2\Psi}{2 \rho^2}\right]\times \nonumber\\
&&\cos\Psi \frac{d\Psi}{d\rho}+\left[\frac{\Delta p}{k_c}+
\frac{\lambda \sin\Psi}{k_c \rho} \right.  \nonumber\\
&& \left.+\frac{c_0^2 \sin\Psi}{2\rho} -\frac{\sin^3\Psi+
2 \sin\Psi \cos^2\Psi}{2\rho^3}\right], \label{3}
\end{eqnarray}
%%%%%%%%%%%%%%%%%%%%%%%%%%%%%%%%%
where $\rho$ is the distance from the symmetric $Z$ axis of rotation,
$\Psi(\rho)$ is the angle made
by the rotational axis and the surface normal of the vesicle, also 
the tangent angle of the contour $Z(\rho)$ 
(see Fig.~\ref {1}).
Let $\sin\Psi=u$, $\rho =x$ \cite{Zheng}, the equation takes the 
following form
\begin{eqnarray}
&&\frac{\Delta p}{k_c}+[\frac{1}{x^3}-\frac{1}{x} \left(\frac{\lambda}{k_c}+
\frac{c_0^2}{2}\right)] u- \frac{u^3}{2x^3} 
-\left(\frac{1}{x^2}+\frac{\lambda}{k_c}+\frac{c_0^2}{2}\right) u' 
\nonumber\\
&&+\frac{3}{2} \frac{u^2 u'}{x^2}-\frac{3u u'^2}{2x}+\frac{2 c_0 u u'}{x}+
\frac{u'^3}{2}+\frac{2(1-u^2)u''}{x} \nonumber\\
&&-uu'u''+(1-u^2)u'''=0, \label{4}
\end{eqnarray}
where prime means derivative of $u(x)$ with respect to x.
It is obvious that a sphere with radius $r_0$ is always a solution of
 Eq.~(\ref{4}) ($u=x/r_0$), 
and its pressure difference $\Delta p \equiv \Delta p_0$ must obey
\begin{eqnarray}
\Delta p_0 r_0^3 -2\lambda r_0^2+k_c c_0 r_0 (2-c_0 r_0)=0. \label{5}
\end{eqnarray}
Jenkins \cite{Jenkins}, Ou-Yang and Helfrich \cite{$Ou-Yang_1$} have studied 
the stability of the spherical vesicles. Now we
calculate the first order pressure perturbation contribution to $u$ using 
the Eq.~(\ref{4}).
When the pressure difference $\Delta p$ is slightly deviated from its 
equilibrium value $\Delta p_0$, i.e. , $\Delta p =\Delta p_0 +
 \delta  \Delta p$,  
 Eq.~(\ref{5}) is no longer satisfied, what will be the shape of the 
vesicle? Here
we give an answer to this question with our detailed calculation.

By expanding $u$ as $u=u_0+u_1+u_2+...$ with $u_1\sim \delta \Delta p$, $
u_2\sim (\delta \Delta p)^2$, ... , while $u_0$ takes $x/r_0$, we find $u_1$ to satisfy the 
following linear third-order differential 
equation
%%%%%%%%%%%%%%%%%%%%%%%%
\begin{eqnarray}
&&\left(1-\frac{x^2}{r_0^2}\right) u_1''' + \left(\frac{2}{x}-
\frac{3 x}{r_0^2}\right)u_1'' \nonumber\\
&&-\left(\frac{1}{x^2}+\frac{\lambda}{k_c}+ %\nonumber\\
\frac{c_0^2}{2}-\frac{2 c_0}{r_0}\right)u_1' \nonumber \\
&&+\left[\frac{1}{x^3}-\frac{1}{x} \left(\frac{\lambda}{k_c}+
\frac{c_0^2}{2}+  
\frac{2 c_0}{r_0}\right)\right] u_1+\frac{\delta \Delta p}{k_c}=0, 
\label{6}
\end{eqnarray}
%%%%%%%%%%%%%%%%%%%%%%%%%%
where $\delta \Delta p $ is the pressure perturbation.
After introducing a new variable $\xi = x/r_0$, i.e., $u_0$ takes $\xi$, the above perturbation 
equation becomes
%%%%%%%%%%%%%%%%%%%%%%%%%%
\begin{eqnarray}
&(&1-\xi^2)u_1'''+(\frac{2}{\xi}-3\xi)u_1''-(\frac{1}{\xi^2}+b)u_1' 
\nonumber\\
&&+(\frac{1}{\xi^3}-\frac{b}{\xi})u_1+p=0, \label{7}              
\end{eqnarray}
%%%%%%%%%%%%%%%%%%%%%%%%
where two dimensionless parameters are:
$b\equiv r_0^2 \lambda/k_c+c_0^2 r_0^2/2-2 c_0 r_0$,
  $p\equiv r_0^3 \delta \Delta p/k_c $, and now prime means derivative of 
  $u(\xi)$ with respect to $\xi$.
It is obvious that $u_1=(p/2b)\xi$ is a particular solution.
Here we suppose that $b \ne 0 $. For the case of $b=0$, we discuss
 it 
in the next section.
Then the first step for solving Eq.~(\ref{7}) is to solve the following homogeneous equation
%%%%%%%%%%%%%%%%%
\begin{eqnarray}
&(&1-\xi^2) u_1'''+(\frac{2}{\xi}-3 \xi)u_1''-(\frac{1}{\xi^2}+b)u_1' 
\nonumber\\
&&+\left(\frac{1}{\xi^3}-\frac{b}{\xi}\right)u_1=0. \label{8}
\end{eqnarray}
%%%%%%%%%%%%%%%%%%%%%%%%%%%%%%%
It is lucky that $u_1=\xi^{-1}$ is a particular solution of this homogeneous equation. 
We denote it as
$u_{10}=\xi^{-1}$, 
and let \cite{Murphy}
\begin{eqnarray}
u_1=\frac{1}{\xi} \int z(\xi) d\xi,\label{9}
\end{eqnarray}
Eq.~(\ref{8}) is thus reduced into the following second-order homogeneous 
differential equation
\begin{eqnarray}
\xi^2 (1-\xi^2)z''-\xi z'+(1-b \xi^2)z=0. \label{10}
\end{eqnarray}
To solve it, we refer to the following standard form of the equation
\cite{$hand_book$}
\begin{eqnarray}
x^2 (ax^n-1) y_{xx}^{''}+ x(atx^n+q)y_{x}^{'}+ (arx^n+s)y=0. \nonumber
\end{eqnarray} 
There are four roots $A_1$, $A_2$ and $B_1$, $B_2$ of the quadratic equations: 
%\begin{eqnarray}
$A^2-(q+1)A-s=0$ and  $B^2-(t-1)B+r=0$. Then four parameters c, $\alpha$,
$\beta$, and $\gamma$ can be defined by the relations: $c=A_1$, $\alpha=\frac{A_1+B_1}{n}$, $\beta=\frac{A_1+B_2}{n}$, $\gamma=\frac{A_1-A_2}{n}+1$ and the         solution
of the standard equation has the form $y=x^c\omega(ax^n)$, where 
$\omega(\xi)$
is the general solution of the hypergeometric equation 
\begin{eqnarray}
\xi (\xi - 1 ) \omega_{\xi \xi}^{''} + [(\alpha+ \beta+1)-\gamma] \omega_\xi^{'}+ \alpha \beta \omega =0  \nonumber. 
\end{eqnarray}

%%%%%%%%%%%%%%%%%%%%%%%%%%%%%%%%%%%%%%%%%%%%%%%%
In the case of Eq.~(\ref{10}), we have $a=1$, $n=2$, $t=0$, $q=1$, $r=b$ and $s=-1$.
 Two quadratic algebra 
equations associated with Eq.~(\ref{10})  are introduced:
%\begin{eqnarray}
$A^2-2 A+1=0$ and $B^2+B+b=0$.  %\ nonumber\\
%\end{eqnarray}
Their roots are:
%\begin{eqnarray}
$A_1=A_2=1$, and $B_1$, $B_2=(-1\pm \sqrt{1-4b})/2$.
%\end{eqnarray}
From them three characterized numbers can be calculated :
\begin{eqnarray}
\alpha&=&\frac{A_1+B_1}{2}=\frac{1+\sqrt{1-4b}}{4}, \nonumber\\
\beta&=&\frac{A_1+B_2}{2}=\frac{1-\sqrt{1-4b}}{4}, \nonumber\\
\gamma&=&1.                \label{11}
\end{eqnarray}
Then we have
\begin{eqnarray}
z(\xi)=\xi \omega(\xi^2), \label{12}
\end{eqnarray}
where $\omega(\eta)$, $\eta \equiv \xi^2$, is the solution of Gauss hypergeometric equation:
\begin{eqnarray}
\eta(\eta-1)\omega''+ [(\alpha+ \beta + 1) \eta - \gamma] \omega'+
\alpha\beta\omega=0. \label{13}
\end{eqnarray}
At once, let its solution $\omega_1$ take the hypergeometric function 
\cite{$hand_book$} as
\begin{eqnarray}
\omega_1=F(\alpha,\beta,\gamma,\eta). \label{14}
\end{eqnarray}
Because $\gamma=1$ is an integral, we have the second solution of 
Eq.~(\ref{13}) \cite{Whittaker} 
%%%%%%%%%%%%%%%%%%%%%%%%%%%%%%%%%
\begin{eqnarray}
\omega_2(\eta)&=&F(\alpha,\beta,\gamma,\eta) \ln(\eta) \nonumber\\
&&+\sum_{k=0}^{\infty}\frac{(\alpha)_k (\beta)_k}{(k!)^2}\eta^k 
\{\phi(\alpha+k) \nonumber\\
&&+\phi(\beta+k)-2\phi(1+k) \},  \label{15}
\end{eqnarray}
%%%%%%%%%%%%%%%%%%%%%%%%%%%%%%%%%%
%%%%%%%%%%%%%%%%%%%%%%%%%%%%%%%%%%
where $(\lambda)_0=1$, $(\lambda)_n=\lambda(\lambda+1)...(\lambda+n-1)= 
%\nonumber\\
\frac{\Gamma(\lambda+n)}{\Gamma(\lambda)}$, $(n \geq 1)$, 
$\phi(\tau)=\frac{d \ln(\Gamma(\tau))}{d\tau}$, and  
$\Gamma(\tau)$ is Euler Gamma function. Finally, we find two 
solutions of Eq.~(\ref{10})
%%%%%%%%%%%%%%%%%%%%%%
\begin{eqnarray}
z_1&=&\xi \omega_1(\eta)=\xi \omega_1(\xi^2)=\xi F(\alpha,\beta,1,\xi^2),
\nonumber\\
z_2&=&\xi \omega_2(\eta)=\xi \omega_2(\xi^2)=\xi F(\alpha,\beta,1,\xi^2)\ln\xi^2
+\nonumber\\
   &&\sum_{k=0}^{\infty}\frac{(\alpha)_k(\beta)_k}{(k!)^2}\eta^{2k+1} 
   \{\phi(\alpha+k) \nonumber\\
   &&+\phi(\beta+k)-2\phi(1+k)\}, \label{16}
\end{eqnarray}
%%%%%%%%%%%%%%%
and two solutions of Eq.~(\ref{8}) $u_{11}=\frac{1}{2\xi}\int\omega_1(\xi^2)
d\xi^2$,
  $u_{12}=\frac{1}{2\xi}\int\omega_2(\xi^2)d\xi^2$.
  To obtain the detailed forms of $u_{11}$ and $u_{12}$ two integrals,   
%\begin{eqnarray}
$\int F(\alpha,\beta,\gamma,\eta)d\eta$ and $\int F(\alpha,\beta,\gamma,\eta)
\ln\eta d\eta$,  need to calculate.  % \nonumber\\
%\end{eqnarray}
We have (see the details in appendix)
%%%%%%%%%%%%%%%%%%%%%
\begin{eqnarray}
&&\int F(\alpha,\beta,\gamma,\eta)d\eta \nonumber\\
&&=\frac{1}{\alpha+\beta-\alpha \beta-1}F(\alpha,\beta,\gamma,\eta)
\left[(\alpha+\beta-1)\eta+1-\gamma\right] \nonumber \\ 
&&+\eta(\eta-1)F'(\alpha,\beta,\gamma,\eta), \label{17}
\end{eqnarray}
%%%%%%%%%%%%%%%%%%%%%%
and, due to $\alpha+\beta-\alpha \beta-1 \neq 0$ 
and $\gamma=1$, 
%%%%%%%%%%%%%%%%%%%%%%% 
\begin{eqnarray}
 &\int& \ln \eta F(\alpha,\beta,\gamma,\eta)d\eta  \nonumber\\
 &&=\frac{1}{\alpha+\beta-\alpha \beta-1}\{F'(\alpha,\beta,\gamma,\eta)
 \eta(\eta-1)\ln\eta \nonumber\\
&&+(\alpha+\beta-1)\eta F(\alpha,\beta,1,\eta)\ln\eta  
+(1-\eta)F(\alpha,\beta,1,\eta)\} \nonumber \\
&&+\frac{2-\alpha-\beta}{(\alpha+\beta-\alpha\beta-1)^2} %\nonumber\\
\{\eta F(\alpha,\beta,1,\eta)(\alpha+\beta-1) \nonumber\\
&&+\eta(\eta-1)F'(\alpha,\beta,1,\eta)\}, \label{18}
\end{eqnarray}
%%%%%%%%%%%%%%%%%%%%%%%%
where $F'(\alpha,\beta,1,\eta)=\frac{d}{d\eta}F(\alpha,\beta,1,\eta)=
\alpha\beta F(\alpha+1,\beta+1,2,\eta)$.  
Now, we can write down three independent solutions of Eq.~(\ref{8}): 

\begin{eqnarray}
u_{10}&=&\frac{1}{\xi}, \nonumber\\
u_{11}&=&\frac{1}{\xi}\int\xi F(\alpha,\beta,1,\xi^2) d\xi \nonumber\\
&=&\frac{1}{2\xi(\alpha+\beta-\alpha \beta-1)}\{F(\alpha,\beta,1,\xi^2)
%\nonumber\\
(\alpha+\beta-1)\xi^2 \nonumber\\
&&+\xi^2(\xi^2-1) \alpha\beta F(\alpha+1,\beta+1,2,\xi^2)\}, \nonumber \\ 
u_{12}&=&\frac{1}{\xi}\int \xi \ln\xi F(\alpha,\beta,\gamma,\xi^2)d\xi
\nonumber \\
&&+\frac{1}{\xi} \int \sum_{k=0}^{\infty}\frac{(\alpha)_k(\beta)_k}{(k!)^2}
\xi^{2k+1} \{\phi(\alpha+k) \nonumber\\
&&+\phi(\beta+k)-2\phi(1+k)\} d\xi \nonumber\\
&=&\frac{1}{2\xi}\{\frac{1}{\alpha+\beta-\alpha \beta-1}  \nonumber\\
&&[\alpha\beta F(\alpha+1,\beta+1,2,\xi^2)\xi^2 (\xi^2-1) \ln\xi^2 \nonumber\\
&&+(\alpha+\beta-1)\xi^2 F(\alpha,\beta,1,\xi^2) \ln\xi^2+ \nonumber\\
&&(1-\xi^2)F(\alpha,\beta,1,\xi^2)]+\frac{2-\alpha-\beta}{(\alpha+\beta-
\alpha\beta-1)^2}\times \nonumber\\ 
&&[\xi^2 F(\alpha,\beta,1,\xi^2)(\alpha+\beta-1)  \nonumber\\
&&+\xi^2(\xi^2-1) \alpha\beta F(\alpha+1,\beta+1,2,\xi^2)]\} \nonumber\\
&&+\sum_{k=0}^{\infty}\frac{(\alpha)_k(\beta)_k}{(k!)^2}\xi^{2k+1} 
\{\phi(\alpha+k)+ 
 \phi(\beta+k) \nonumber\\
&&-2\phi(1+k)\}.  \label{19}
\end{eqnarray}
From Eq.~(\ref{7}),  Eq.~(\ref{8}) and Eq.~(\ref{19})
we get the general solution of Eq.~(\ref{4}) as 
\begin{eqnarray}
u=u_0+p c_{10} \frac{1}{\xi}+p c_{11} u_{11} +p c_{12} u_{12}+
p \frac{1}{2b}\xi, \label{20}
\end{eqnarray}
where $c_{10}$, $c_{11}$, and $c_{12}$ are three integral constants. 

In general, the hypergeometric function can be expressed as the 
series 
\begin{eqnarray}
F(\alpha,\beta,\gamma,\eta)=1+\sum_{k=0}^{\infty}
\frac{(\alpha)_k(\beta)_k}{(\gamma)_k} \frac{\eta^k}{k!},\label{21} 
\end{eqnarray}
where $(\lambda)_0=1, (\lambda)_n=\lambda(\lambda+1)...(\lambda+n-1)$.
$F(\alpha, \beta, \gamma, \eta)$ is a fortiori convergent for $|\eta|<1$. In the present case $\gamma=1$,
 $\alpha=(1+\sqrt{1-4b})/4$ and 
$\beta=(1-\sqrt{1-4b})/4$, the condition $Re(\gamma-\alpha-\beta)=
\frac{1}{2}>0$
is satisfied, so $F(\alpha,\beta,1,\eta)$ is conditional convergent while 
$\eta=1$,  
and $F(\alpha,\beta,\gamma,1)=\frac{\Gamma(\gamma)\Gamma(\gamma-\alpha-\beta)}
{\Gamma(\gamma-\alpha)\Gamma(\gamma-\beta)}$.
But this case is not true to $F(\alpha+1,\beta+1,2,\eta)$ when $\eta=1$, 
and the
analytical continuation for it gives the complex result. So we limit our discussion 
within $0 \leq\xi<1$ 
to avoid the divergence and imaginary number. From another point of view, if we
 consider
that $\xi$ can be or be more than unit, that three integral constants 
$c_{10}$, $c_{11}$ and $c_{12}$ in Eq.~(\ref{19}) must be
 zero, 
so the solution of Eq.~(\ref{4}) will be $u_0$ (sphere solution)
 adding 
that
particular one, i.e.,
\begin{eqnarray}
u=u_0 + \frac{p}{2 b} \xi, \nonumber
\end{eqnarray}
the sphere still keeps its shape under a 
pressure perturbation
varying its size only. After obtaining the total solution of 
Eq.~(\ref{6}), using the integration
 
\begin{eqnarray}
Z = -\int \tan \Psi d \rho,  \label{22}
\end{eqnarray}
we can give the contour
of the axial symmetry shape of vesicle with deformation. 
Our numerical calculation about Gauss hypergeometric function
is worked out with the famous software, {\it Mathematica}. In the following sections
some cases are shown for details.

%%%%%%%%%%%%%begin of special case%%%%%%%%%%%%%%%%%%%
\section{special case }
First we discuss the special case of $b=0$ in the  
Eq.~(\ref{7}). That is to solve the following equation
\begin{eqnarray}
(1-\xi^2)u'''+(\frac{2}{\xi}-3\xi)u''-\frac{1}{\xi^2}u'+\frac{1}{\xi^3}u+p=0. \label{23}
\end{eqnarray}
It is convenient to check that $u=p/2\xi\log{\xi}$ is a particular solution. So the
task remained is to solve the associated homogeneous third order equation
\begin{eqnarray}
(1-\xi^2)u'''+(\frac{2}{\xi}-3\xi)u''-\frac{1}{\xi^2}u'+\frac{1}{\xi^3}u=0. \label{24}
\end{eqnarray}
It is easy to verify that $u_1=\xi$, $u_2=1/\xi$ are two independent solutions.
According to the handbook \cite{$hand_book$}, the general solution of 
Eq.~(\ref{24})
is written as 
\begin{eqnarray}
u=c_1u_{10}+c_2u_2+c_3(u_2 \int{u_1\theta dx}-u_1\int{u_2 \theta dx}) \label{25} 
\end{eqnarray}
where
\begin{eqnarray}
\theta=\exp[\int(f_2/f_3)dx](u_1u_2'-u_1'u_2)^{-2}, \nonumber
\end{eqnarray}
\begin{eqnarray}
f_2=2/\xi-3\xi,  \quad  and \quad f_3=1-\xi^2.  \nonumber 
\end{eqnarray}
%and
%\begin{eqnarray}
%f_3=1-\xi^2.  \nonumber
%\end{eqnarray}
%$f_2=2/\xi-3\xi$, and $f_3=1-\xi^2$.
The final result is 
\begin{eqnarray}
u&=&\xi+c_1\xi+\frac{c_2}{\xi}+c_3(\xi\log{\frac{1+\sqrt{1-\xi^2}}{\xi}})  \nonumber \\
&&+\frac{p}{2}\xi\log{\xi}. \label{26} 
\end{eqnarray}
If \qquad $c_2=c_3=0$, we have 
\begin{eqnarray}
u=\xi+c_1\xi+\frac{p}{2}\xi \log{\xi}. \label{27} 
\end{eqnarray}
Eq.~(\ref{27}) satisfies the general shape equation Eq.~(\ref{4}). It can 
give the biconcave shape \cite{Naito} and 
many interesting shapes described as have
been shown in \cite{QuanHui}. 

%%%%%%%%%%%%%%%%%%%end of special case %%%%%%%%%%%%%%%%%%%%%%%
\section{Biconcave and peanut shape}
With the help of microscopes people have made detailed observation
of RBC which can assume various shapes \cite{Bessis}. 
Generally, RBC takes biconcave disk shape (in blood capillary
 it has very large deformation \cite{Branemark}), while some pathologic
 cells take other abnormality (such as in sickle cell disease).
 If RBC is subjected to different environment, various pH
 values for example, it will make some considerable deformations, 
such as, Cup (Stomatocyte), Bell (Codocyte), Sea urchin
 (Echinocyte) {\it et al.}. Moreover, beginning with biconcave
 lipsome vesicle Hotani present many beautiful transformations
 among which the peanut shape was included 
 \cite{Hotani}. Surely the biconcave shape attracts many 
attention.
In this section we use the general theory of the section II to show 
that spherical
vesicle can transform into biconcave shape via pressure perturbation. 
Because $\Psi=0$ at $\rho=0$ and $\Psi=\pi/2$ at $\rho=\rho_m$ (see
 Fig.~\ref{1}),  
there are two boundary conditions at $\rho=0$ and $\rho=\rho_m$, i.e., 
$\xi=0$ and $\xi=\xi_m$
\begin{eqnarray}
u|_{\xi=0}=0; \qquad  \qquad    u|_{\xi=\xi_m}=1, \label{28}
\end{eqnarray}
where $\xi_m=\rho_m/r_0$.
Thus there are two relations between the three coefficients $c_{10}$, $c_{11}$ and  
$c_{12}$ as  
\begin{eqnarray}
c_{10}&=&\frac{-c_{12}}{2(\alpha+\beta-\alpha\beta-1)}, \nonumber\\
c_{11}&=&\{c_{12} \{\frac{2-\alpha-\beta}{(\alpha+\beta-\alpha\beta-1)}
\frac{1}{2(\alpha+\beta-\alpha\beta-1 )}\times\nonumber\\
&&\frac{\Gamma(\gamma)\Gamma(\gamma-\alpha-\beta)}{\Gamma(\gamma-\alpha)
\Gamma(\gamma-\beta)}- %\nonumber\\
\frac{1}{2(\alpha+\beta-\alpha\beta-1 )}\}+\frac{1}{2b}\}\times \nonumber\\
&& \frac{ 2(\alpha+\beta-\alpha\beta-1)\Gamma(\gamma-\alpha)
\Gamma(\gamma-\beta)}
{\Gamma(\gamma)\Gamma(\gamma-\alpha-\beta)(\alpha+\beta-1)}. \nonumber\\
 \label{29}
\end{eqnarray}
So far, in the above solution there is still one relation to be determined.
 We can 
calculate
it via using the conservation of  area of sphere.
For a rotational surface the area element can be expressed as 
\begin{eqnarray}
ds&=&2\pi x(1-u^2)^{-1/2}dx.  \nonumber
\end{eqnarray}
Using reduced dimensionless area $s_r$ defined by $ds=4\pi r_{0}^2 ds_r$,
we have  
 $ds_r=\frac{1}{2}\xi (1-u^2)^{-1/2}d\xi$, 
and another boundary condition to take account of the area 
conservation 
\begin{eqnarray}
\int_0 ^{\xi_m} \xi (1-u^2)^{-1/2}d\xi=1.  \label{30}
\end{eqnarray}
%%%%%%%%%%%%%%%%%%%%%%%%%%%%%%%%%%%%%%%%%%%%%%%%%%%
Basing on Eq.~(\ref{20}), Eq.~(\ref{29}) and Eq.~(\ref{30}),
the values of $c_{10}$, $c_{11}$ and $c_{12}$ can be fixed 
completely. 
So the shape is also determined with Eq.~(\ref{22}). But 
in practical computation the
value of $\xi_m$ in Eq.~(\ref{30}) can not be  directly predicted. 
Therefore, in our computation, we treat it as a 
free boundary problem, i.e., we let $\xi_m$ takes
a series of value, correspondingly, a series of deformation
of the original spherical vesicle and their total energy
can be calculated \cite{$Total_energy$}. The less total energy, 
the easier the deformation can be observed in practice (experiments).
 From the Eq.~(\ref{20}) the different shapes can be
got via Eq.~(\ref{22}). We show our numerical results in Fig.~\ref{2},
Fig.~\ref{3} and Fig.~\ref{4}. In fact, in Fig.~\ref{2}
only the top right quadrant contours of vesicle are shown.
In order to obtain the total vesicle surface two operations
need to be done in succession: at first, rotate it with the
$Z$ axis, this give the upper part of the surface; second,
take mirror reflection. In order to describe the parameters
 corresponding
to the curves in Figs.~\ref{2}-6, we define a new parameter $v$ 
as $v$ $\equiv$ $V/V_0$ where $V$ is the volume
of vesicle and 
$V_0$ is the volume of original spherical vesicle.
The pressure corresponding to each curve is defined as
$\frac{\Delta p-\Delta p_0}{\Delta p_0}$, $\Delta p$ is the pressure
 related
to the deformed vesicle while $\Delta p_0$ is the pressure difference 
of the original spherical vesicle.
%%%%%%%%%%%%%%%%%%%%%%%%%%%%%%%%%%%%%%%%%%%%%%%%%%%%%%%%%
The shapes given in Fig.~\ref{2} are obtained at the constant 
spontaneous curvature and surface tension, i.e., $c_0r_0$ 
=-2.4 and $\lambda r_{0}^2/k_c=0.01$. The values of 
($\frac{\Delta p-\Delta p_0}{\Delta p_0}$, $v$) are calculated as ($-10^{-6}$, 0.99), (0.01, 0.3), (0.01, 0.5),
 (0.05, 0.3), (0.01, 0.62) for
 curves 2-6, respectively. The curve 1 is the initial shape
 of sphere. Two three-dimensional shapes of biconcave and peanut 
are shown in Fig.~\ref{3} and Fig.~\ref{4}, respectively. 
In principle, energy corresponding to every curve can be calculated
 with the formula in \cite{$Total_energy$}.
The above calculated biconcave shape can be simulated 
on the basis of the elasticity theory of Helfrich spontaneous
 curvature model by {\it{Surface
Evolver}} {\cite{Brakke}}.
The simulated
results, corresponding to biconcave and peanut shape, 
are shown in Fig.~\ref{5} and Fig.~\ref{6}, respectively, 
in which the $A/A_0$ gives the ratio of surface area of vesicle to that of 
the initial spherical vesicle.   
Both they 
are observed in blood cell \cite{Bessis} and lipsome vesicles 
\cite{Hotani}.
Furthermore, We turn to analysis the behavior of deformation near rotational axis.
% \section{behavior of deformation near rotational axis}
 After obtaining the general solution Eq.~(\ref{20}),
 the behavior of deformation near 
 the 
 rotational axis can be got by expanding it to first order of $p$ as
 \begin{eqnarray}
 u&=&\xi+\frac{1}{2}p c_{11}\xi+p c_{12} (\ln\xi+\frac{1}{2}+
 \frac{\frac{1}{2}\alpha\beta}{\alpha+\beta-\alpha\beta-1}) \xi+
 \frac{p}{2 b} \xi \nonumber\\
 &=&A \xi +B \xi \ln\xi.  \label{31}
 \end{eqnarray}
 This equation describes the shape near polar points, and the equation
 is very similar to the biconcave shape of red blood
 cell \cite{Naito}. In this
 equation there is no any singular point in the region $0 \leq \xi<1$.
%%%%%%%%%%%%begin of Myelin form%%%%%%%%%%%%%%% 
\section{Myelin form}
Myelin form may originate from all blood cells \cite{Bessis}. 
Shadowing technique in microscopy reveals that myelin forms are hollow structures. 
When a RBC is in aging, aged and damaged states, it gives rise to 
large myelin forms which may take various types: filaments, 
beads or strings of beads. These
filaments, which are easily seen with phase contrast microscopy,
 may remain
 attached to the surface of the RBC at one end.
Here we choose the so called Delaunay's surface solution \cite{Naito}
 from
 the general solution of the perturbation 
Eq.~(\ref {6}), which can 
give a type of myelin form generating from the vesicle.
 In the book of Bessis \cite {Bessis}, there is a photo of this
kind of Medusa head.
  Now let us first discuss the general solution Eq.~(\ref{20}) 
in the case of $c_{11}=c_{12}=0$:
\begin{eqnarray}
u=(1+\frac{p}{2b})\xi+c_{10}/\xi. \label{32}
\end{eqnarray}
Substituting it into
 the general shape
equation Eq.~(\ref{4}) yields two relations: 
$1+p/2b=c_0 r_0/2$, and $\Delta p=c_0 \lambda$. It should be
 noticed that
$\Delta p=\Delta p_0+p$. $\Delta p_0$ can be solved from Eq.~(\ref{5}),
 then we find
\begin{eqnarray}
p=(2 c_0-c_{0}^2 r_0) r_0-\frac{2 \lambda r_{0}^2}{k_c}
 +\frac {c_0 \lambda r_{0}^3}{k_c}.
\label{33}
\end{eqnarray}
Considering the mentioned relation $1+p/2b =c_0 r_0$ we have
\begin{eqnarray}
(c_{0}^2 r_0 -2c_0) r_0+\frac{2 \lambda r_{0}^2}{k_c}
 -\frac {c_0 \lambda r_{0}^3}{k_c}=2b -c_0 r_0 b.  \nonumber
\end{eqnarray}
The value of $c_{10}$ in Eq.~(\ref{32}) can be determined via
 boundary condition: 
$u|_{\xi=\xi_m}=1$ ($\xi=\xi_m$, $\Psi=\pi/2$).
It leads to $c_{10}=(\xi_m - (1+\frac{p}{2b})\xi_{m}^2)$  and 
finally we obtained from Eq.~(\ref{32}) the Delaunay's solution
\begin{eqnarray}
u=\frac{c_0 r_0}{2} \xi + \frac{\xi_m - c_0 r_0 \xi_{m}^2 /2}{\xi}.    \label{34}
\end{eqnarray}
With different $c_0 r_0$ and $\xi_m$, the various contours can be
numerically calculated using Eq.~(\ref{22}) (see Fig.~\ref{7}
 and Fig.~\ref{8}). Not total length of filament is drawn in this two figures. In fact, those filaments are swollen distally and our present calculation
can not give an exact description yet. 
%%%%%%%%%%%%end of Myelin form%%%%%%%%%%%%%%%
%%%%%%%%%%%%%%%%%  conclusion and discussions %%%%%%%%%%%%%%%%%%
\section{conclusions and discussions}
Now we give here our main conclusions as what follows: With tangent
 angle perturbation 
approach,
we can calculate the large deformation of spherical vesicle under
 pressure
 perturbation.
From the general perturbation solutions, the biconcave and peanut shapes can
be obtained and a kind of myelin form is shown to be existed.
 Our all calculations are based on the 
elasticity theory of Helfrich spontaneous curvature model and it 
does give good accordance with some complex shapes of lipsome
 vesicles (see photographs in \cite{Hotani})
and RBCs (see photographs in \cite{Bessis})
with  computer simulation \cite{Yan Jie,Bessis,Hotani}.
%%%%%%%%%%%%%%%%%end of discussion%%%%%%%%%%%%%%%%%%%%%%%%%%%%%%%%% 
\acknowledgements
{One of the authors(J. Zhou) thanks  Dr. Y. Zhang, Prof. W. M. Zheng
 and Prof. 
H. W. Peng
for stimulating discussions, especially for the help from
 Dr. H. J. Zhou, 
 Dr. J. Yan and Dr. W. Y. Wang.}

\appendix 
\section*{two integrals relating to Gauss hypergeometric function}
In the appendix we give the derivation of two relating integral about Gauss
 hypergeometric
 function. The main idea and method are to use hypergeometric
 equation itself.
  From the hypergeometric equation Eq.~(\ref{13})
\begin{eqnarray} 
 x~(1-x)\omega '' + \left[\gamma-~(\alpha+\beta+1)x\right]\omega'-
 \alpha \beta \omega=0, \nonumber 
\end{eqnarray}
we have
 \begin{eqnarray}
 \omega=\frac{1}{\alpha \beta}x~(1-x)\omega^{\prime \prime}+
\left[\gamma-~(
 \alpha+\beta+1)\right]\omega^{\prime}, \nonumber
\end{eqnarray} 
and 
\begin{eqnarray}
\int \omega dx = \frac{x~(1-x)\omega^{\prime}}{1-\alpha-\beta+
\alpha \beta}+
\frac{\gamma-1+~(\alpha+\beta+1)x}{1-\alpha-\beta+\alpha \beta}\omega. 
%\nonumber
\end{eqnarray} 
Using the same method and  with the key point $\gamma=1$, through a
 lengthy derivation,
we obtain 
\begin{eqnarray}
\int \ln x \omega dx &=& \frac{1}{\alpha+\beta+1}\times \nonumber \\
&&[\omega^{\prime}x~(x-1)\ln x+~(\alpha+\beta-1)x\omega\ln x \nonumber\\
&&+(1-x)\omega] %\nonumber \\
+\frac{2-\alpha-\beta}{\left(\alpha+\beta-\alpha\beta-1\right)^2}\times 
\nonumber\\
&&[x\omega~(\alpha+\beta-1)+x(x-1)\omega^{\prime}]. \nonumber\\ 
\end{eqnarray}
%%%%%%%%%%%%%%%%%%%%%%
%%%%%%%%%%%%%%%%%%%%%%%%%%%%%%%%%%%%%%%%%%%%%%%%%

%%%%%%%%%%%%%%%%%%%%%%%%%%%%%%%%%%%%%%%%%%%%%%%%%%%%%%%%%%%%%%%%%%%
%\newpage
\begin{figure}
\centerline{\epsfxsize=4cm \epsfbox{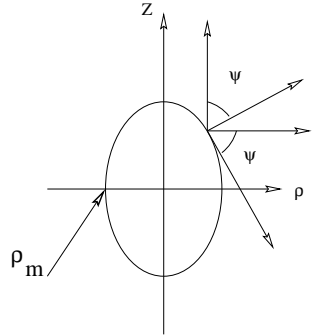}}
\caption{Convection of angle $\Psi$ for the contour of axial symmetry 
vesicle. At the north pole the $\Psi$ value takes zero. $\rho_m$ gives
the boundary position of $\rho$.
}
\label{fig1}
\end{figure}
%%%%%%%%%%%%%%%%%%%%%
%\newpage
\begin{figure}
\centerline{\epsfxsize=4cm \epsfbox{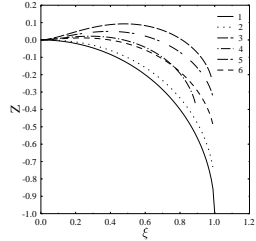}}
\caption{The numerical results from the solution of Eq.~(\ref{20}). Only the
 top right quadrant contours are shown. Curve 1 corresponds 
to spherical vesicle, while curves 2-6 show much swelling or oblate 
biconcave shapes.
 The energy of each vesicle can be calculated 
with the formula given in [18].}
\label{fig2}
\end{figure}
%%%%%%%%%%%%%%%%%%%

%\newpage
\begin{figure}
\centerline{\epsfxsize=4cm \epsfbox{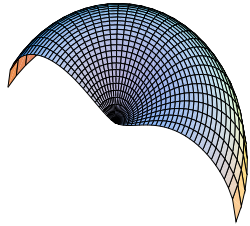}}
\caption{
Biconcave shape in three-dimensional (half of upper part is shown) with
$\xi_m=0.9999$,
$v=0.1$ (the ratio of volume of the deformed vesicle to that of
 the initial sphere),
 $\lambda r_{0}^2/k_c=0.01$, $c_0r_0=-3.0$. When the parameters are
adjusted, the other biconcave shapes can be formed,
 much oblate or
much swelling (see, Fig.~\ref{2}).
}
\label{fig3}
\end{figure}
%%%%%%%%%%%%%%%%%%%%%

%\newpage
\begin{figure}
\centerline{\epsfxsize=4cm \epsfbox{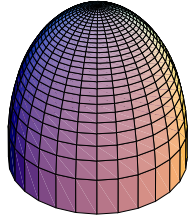}}
\vspace{0.6cm}
\caption{ 
Numerical result of upper-half peanut shape in three-dimensional
 with $\xi_m=0.8$,
 $v=0.4$, $\lambda r_0^{2}/k_c=0.01$, and 
$c_0r_0=0.01$.
}
\label{fig4}
\end{figure}
%%%%%%%%%%%%%%%%%%%%%%%
%\newpage
\begin{figure}
\centerline{\epsfxsize=3.5cm \epsfbox{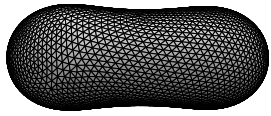}}
\vspace*{0.5cm}
\caption{
Biconcave shape in three-dimensional simulated with {\it Surface Evolver}.
 The parameters take 
$H_0$=-1.38, $k_c$=1, $A/A_0$=0.96 (the ratio of surface area of vesicle and  
that of the initial sphere), $v$=0.92,  
and $\lambda r_{0}^2/k_c$=0.48.
}
\label{fig5}
\end{figure}
%%%%%%%%%%%%%%%%%%%%%%%%
%\newpage
\begin{figure}
\centerline{\epsfxsize=3.5cm \epsfbox{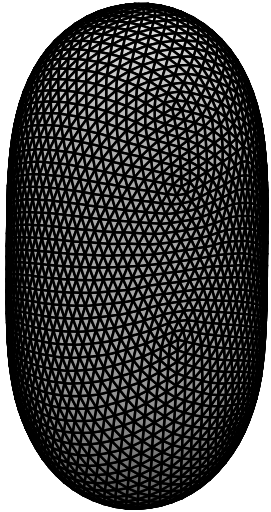}}
\vspace{0.6cm}
\caption{ 
Peanut shape in three-dimensional simulated with {\it Surface Evolver}.
 The using parameters are  
$c_0 r_0$=1.38, $k_c$=1, $A/A_0$=0.83 (the ratio between surface area of 
vesicle and that of the initial sphere), $v$=0.79,
 and $\lambda r_{0}^2/k_c$=0.48.
}
\label{fig6}
\end{figure}
%%%%%%%%%%%%%%%%%%%%%
%\newpage
\begin{figure}
\centerline{\epsfxsize=4cm \epsfbox{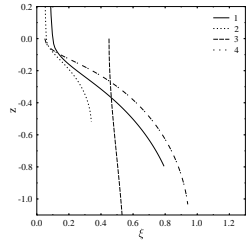}}
%\vspace{0.6cm}
\caption{
The numerical results corresponding to Eq.~(\ref{23}). Only the top right
quadrant contours are shown. The parameters corresponds to curve 4
 (contour in Fig.~\ref{8})
 are: $c_0 r_0=2$, $\xi_m=0.9$.
} 
\label{fig7}
\end{figure}
%%%%%%%%%%%%%%%%%%%%%
%\newpage
\begin{figure}
\centerline{\epsfxsize=4cm \epsfbox{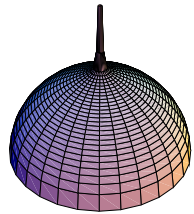}}
\vspace{0.6cm}
\caption{The calculated shape of upper-half of the myelin
 form of spherical
vesicle in three-dimensional with
$c_0r_0=2$, and $\xi_m=0.9$.}
\label{fig8}
\end{figure}
\end{document}